\RequirePackage{fix-cm}
\documentclass{svjour3}                     

\smartqed  
\usepackage{amsmath,amssymb}
\usepackage{amsfonts}
\usepackage{mathtools,tikz}
\usepackage{graphicx}
\usepackage{epstopdf, epsfig}
\usepackage{scalerel,stackengine}
\usepackage{subfig}
\usepackage{floatrow}
\usepackage{float}
\usepackage{mwe}
\usepackage{bm}
\usepackage{onimage}
\usepackage{tikz}
\usetikzlibrary{positioning}
\usepackage[ruled,vlined]{algorithm2e}
\usepackage{appendix}
\usepackage[export]{adjustbox}
\usepackage{calc}
\usepackage{amssymb}

\stackMath
\newcommand\reallywidehat[1]{%
\savestack{\tmpbox}{\stretchto{%
  \scaleto{%
    \scalerel*[\widthof{\ensuremath{#1}}]{\kern-.6pt\bigwedge\kern-.6pt}%
    {\rule[-\textheight/2]{1ex}{\textheight}}
  }{\textheight}%
}{0.5ex}}%
\stackon[1pt]{#1}{\tmpbox}%
}

\DeclareTextFontCommand\textsfi{\usefont{OT1}{cmss}{m}{sl}}
\DeclareMathAlphabet\mathsfi            {OT1}{cmss}{m}{sl}
\DeclareTextFontCommand\textsfb{\usefont{OT1}{cmss}{bx}{n}}
\DeclareMathAlphabet\mathsfb            {OT1}{cmss}{bx}{n}
\DeclareTextFontCommand\textsfbi{\usefont{OT1}{cmss}{m}{sl}}
\DeclareMathAlphabet\mathsfbi            {OT1}{cmss}{m}{sl}
  
\newcommand{\mat}[1]{{\bm{#1}}}
\newcommand{\mA}{\mat{A}}

\newcommand{\mH}{\mat{H}}

\newcommand{\mI}{\mat{I}}

\newcommand{\mT}{\mat{T}}
\newcommand{\Ttilde}{\widetilde{\mat{T}}}

\newcommand{\vect}[1]{{\bm{#1}}}
\newcommand{\vf}{\vect{f}}

\newcommand{\vh}{\vect{h}}
\newcommand{\vq}{\vect{q}}
\newcommand{\vQ}{\vect{Q}}
\newcommand{\vu}{\vect{u}}
\newcommand{\vv}{\vect{v}}

\newcommand{\vw}{\vect{w}}
\newcommand{\vx}{\vect{x}}

\newcommand{\vz}{\vect{z}}

\newcommand{\tvv}{\widetilde{\vv}}
\newcommand{\tvu}{\widetilde{\vu}}

\newcommand{\norm}[1]{\left\lVert#1\right\rVert}
\newcommand{\abs}[1]{\left\vert#1\right\vert}

\begin{document}

\title{A computationally efficient approach for the removal of the phase shift singularity in harmonic resolvent analysis}


\titlerunning{Removal of the singularity in the harmonic resolvent analysis}        

\author{Alberto Padovan         \and
        Clarence W. Rowley 
}


\institute{Alberto Padovan \at
              Department of Mechanical and Aerospace Engineering, Princeton University, NJ 08544, USA \\
              \email{apadovan@princeton.edu}    
           \and
           Clarence W. Rowley \at
              Department of Mechanical and Aerospace Engineering, Princeton University, NJ 08544, USA \\
              \email{cwrowley@princeton.edu}
}

\date{Received: date / Accepted: date}

\maketitle

\begin{abstract}
The recently introduced harmonic resolvent framework \cite{padovan2020} is concerned with the study of the input-output dynamics of nonlinear flows in the proximity of a known time-periodic orbit. These dynamics are governed by the harmonic resolvent operator, which is a linear operator in the frequency domain whose singular value decomposition sheds light on the dominant input-output structures of the flow.
Although the harmonic resolvent is a mathematically well-defined operator, the numerical computation of its singular value decomposition requires inverting a matrix that becomes exactly singular as the periodic orbit approaches an exact solution of the nonlinear governing equations. 
The very poor condition properties of this matrix hinder the convergence of classical Krylov solvers, even in the presence of preconditioners, thereby increasing the computational cost required to perform the harmonic resolvent analysis. 
In this paper we show that a suitable augmentation of the (nearly) singular matrix removes the singularity, and we provide a lower bound for the smallest singular value of the augmented matrix. We also show that the desired decomposition of the harmonic resolvent can be computed using the augmented matrix, whose improved condition properties lead to a significant speedup in the convergence of classical iterative solvers. 
We demonstrate this simple, yet effective, computational procedure on the Kuramoto-Sivashinsky equation in the proximity of an unstable time-periodic orbit.

\end{abstract}

\section{Introduction}
\label{sec: intro}

Throughout the years some fluid flows of interest have been studied using a splitting approach, whereby the nonlinear terms in the Navier-Stokes equation are treated as forcing that acts on the linear terms. This approach is particularly convenient since the linear dynamics lend themselves to classical linear analyses that can help uncover some of the fundamental mechanisms behind complicated physical phenomena. 
For instance, the input-output analysis of the linearized Navier-Stokes operator about a steady base flow has helped shed light on some of the energy amplification mechanisms in shear flows of interest. 
In particular, \emph{resolvent analysis} was used to study the response of perturbations to spatio-temporal forcing in linearized channel flow \cite{MRJ2005}. 
A similar approach was implemented in \cite{BJM2010} to indentify the most amplified velocity structures at selected temporal frequency-spatial wavenumber pairs of interest in turbulent pipe flow.   

Recently, this linear input-output framework was extended to analyze the dominant dynamics of perturbations about periodically time-varying base flows \cite{padovan2020}. 
This extended framework, known as \emph{harmonic resolvent analysis}, is based on the singular value decomposition of the harmonic resolvent operator, a frequency-domain linear input-output operator that governs the dynamics of time-periodic perturbations about a time-periodic base flow.
Much like the singular value decomposition of the resolvent operator discussed in \cite{MRJ2005,BJM2010}, the singular value decomposition of the harmonic resolvent provides insight into the dominant input-output structures of the flow.   

While the harmonic resolvent operator is mathematically well-defined, the numerical computation of its singular values requires some care. 
In particular, we will see in section \ref{subsec: comp_proc_chall} that computing the desired decomposition of the harmonic resolvent requires inverting the linearized Navier-Stokes operator evaluated about the periodic base flow. 
It is well known, however, that linearized periodic dynamics are neutrally stable in the direction of a phase shift along the periodic orbit \cite{guckenheimer}. The linearized Navier-Stokes operator will thus have a one-dimensional nullspace along the direction of the phase shift, and this singularity hinders the performance of classical Krylov-based solvers. 
This is especially problematic in large-scale applications, where iterative solvers may be the only computationally feasible algorithms for the solution of linear systems.


We will show in section \ref{sec: remove_sing} that a suitable augmentation of the linearized Navier-Stokes operator allows for the removal of the singularity, and we provide a lower bound for the smallest singular value of the augmented operator. 
We then show that the desired singular value decomposition of the harmonic resolvent operator can be obtained by working with the better conditioned augmented matrix.
In section \ref{sec:KS} we apply this computational procedure to the Kuramoto-Sivashinsky equation in the proximity of an unstable periodic orbit, and we demonstrate the speedup in convergence that is obtained by properly removing the singularity.  

Before moving forward, it is worth mentioning that the procedure we propose shares some similarities with algorithms for the solution of singular linear systems, available in linear algebra packages such as PETSc \cite{petsc}. These usually rely on the knowledge of the nullspace of the linear operator and of its complex conjugate transpose to compute the least squares solution for the linear system at hand. 


\section{Harmonic resolvent formulation}
\subsection{Mathematical formulation}
\label{sec: math_formulation}
In this section we review the harmonic resolvent operator as formulated in \cite{padovan2020}. We reproduce the derivation here for the sake of clarity. 
We consider a nonlinear system with state $\vq(t)$, and dynamics given by
\begin{equation}
\label{eqn: generalSys}
    \frac{\mathrm{d}}{\mathrm{d}t}\vq(t) = \vf\big(\vq(t)\big).
\end{equation}
We then decompose the state about a periodic base flow $\vQ(t)$ with period $T$:
\begin{equation}
\label{eqn: decomp}
    \vq(t) = \vQ(t) + \vq'(t) = \sum_{\omega \in \Omega_b} \hat\vQ_{\omega}e^{i\omega t} + \sum_{\omega \in \Omega}\hat\vq'_{\omega}e^{i\omega t},
\end{equation}
where $\Omega_b \subseteq \Omega \subset \frac{2\pi}{T}\mathbb{Z}$. Here, $\Omega_b$ is the set of frequencies associated with the base flow, while $\Omega$ is the set of frequencies associated with the perturbations that we wish to resolve.
Upon substituting (\ref{eqn: decomp}) into (\ref{eqn: generalSys}) one obtains
\begin{equation}
\label{eqn: timePeriodic_dynamics}
    \frac{\mathrm{d}}{\mathrm{d}t}\vq'(t) = \underbrace{\mathcal{D}_{\vq}\vf\left(\vQ(t)\right)}_{\mA(t)}\vq'(t) + \vh'(t)
\end{equation}
where $\vh'(t)$ contains higher-order terms. 
Formula~(\ref{eqn: timePeriodic_dynamics}) can be written in the frequency domain as
\begin{equation}
  \label{eqn: freqDomain_Equation}
  i\omega \hat\vq'_{\omega} = \sum_{\alpha\in \Omega}
  \hat\mA_{\omega-\alpha} \hat\vq'_{\alpha}
  + \hat\vh'_{\omega} \qquad \forall \omega \in \Omega.
\end{equation}
For ease of notation, let $\hat\vq'$ be the vector of
$\hat\vq'_\omega$ for all frequencies $\omega \in \Omega$, and let
$\hat\vh'$ be defined similarly.  We then define
the operator $\mT$ by
\begin{equation}
  \label{eq:2}
  \big[\mT\hat \vq'\big]_\omega = i\omega \hat \vq_\omega' - \sum_{\alpha\in\Omega}
  \hat \mA_{\omega-\alpha}\hat \vq_\alpha'.
\end{equation}
If the base flow $\vQ(t)$ is an exact solution of (\ref{eqn: generalSys}), then it is straightforward to see that $\mT$ is singular.  In particular, $\vQ(t+\tau)$ is also an exact solution, for any shift~$\tau$, and by differentiating with respect to $\tau$, one can show that $(\mathrm{d}/\mathrm{d}t)\vQ(t)$ exactly satisfies (\ref{eqn: timePeriodic_dynamics}) with $\vh'=0$, and thus the Fourier coefficients of $\mathrm{d}\vQ/\mathrm{d}t$ lie in the nullspace of $\mT$.  That is, the nullspace of $\mT$ is along the direction of a phase shift of the base flow.

On the other hand, if $\vQ(t)$ is an \emph{approximate\/} solution of (\ref{eqn: generalSys}), then $\mT$ is \emph{nearly} singular along the direction of the phase shift.
As we are not usually interested in the trivial phase shift, the harmonic resolvent was defined in \cite{padovan2020} in a way that removes it.
Specifically, letting $\vv$ be the unit-norm vector in the direction of phase shift given by $\reallywidehat{\mathrm{d}\vQ/\mathrm{d}t}$, and letting $\Sigma = \vv^{\perp}$ denote its orthogonal complement, we define the restricted operator
\begin{equation}
\label{eqn: restricted_op}
    \mT\vert_{\Sigma}: \Sigma \rightarrow W_{\Sigma},
\end{equation}
where $W_{\Sigma}$ (the range of $\mT\vert_\Sigma$) is a codimension-1 subspace orthogonal to a unit-norm vector~$\vu$. We will further discuss $\vu$ and its efficient computation in the upcoming sections. 
Notice that restricting the range and domain of the operator $\mT$ is analogous to constructing a Poincar\'e map by reducing the dynamics onto a codimension-1 subspace pierced by the limit cycle \cite{guckenheimer}. 
Finally, the harmonic resolvent on $W_{\Sigma}$ is defined as
\begin{equation}
    \label{eqn: harmonic_res}
    \mH = \left(\mT\vert_{\Sigma} \right)^{-1}.
\end{equation}
Upon the removal of the phase shift direction, formula (\ref{eqn: freqDomain_Equation}) may be written as 
\begin{equation}
    \label{eqn: input_output}
    \mT \hat\vq' = \hat\vh' \iff
    \hat\vq' = \mH \hat\vh'.
\end{equation}

\subsection{Computational procedure and challenges}
\label{subsec: comp_proc_chall}
As mentioned in the introduction, we are interested in computing the singular value decomposition (SVD) of the harmonic resolvent in order to shed light on the dominant input-output structures of the flow in the proximity of the time-periodic base flow. 
However, given the high dimensionality of systems arising from the discretization of partial differential equations, $\mH$ cannot be computed and stored explicitly. Instead, given the sparse operator $\mT$, one can use a randomized SVD algorithm to compute the leading singular values and singular vectors of the harmonic resolvent.
We refer the reader to \cite{rsvd} for a detailed description of the algorithm. For the upcoming discussion, it suffices to point out that the computation of the randomized SVD requires evaluating matrix-vector products of the form $\mH\hat\vh'$ and $\mH^*\hat\vq'$, where $\mH^*$ is the complex conjugate transpose of $\mH$. Specifically, in order to evaluate $\mH\hat\vh'$ we solve the linear system 
\begin{equation}
\label{eqn: linSys_proj}
    \mT \hat\vq' = \left(\mI - \vu\vu^* \right) \hat\vh'
\end{equation}
where $\left(\mI - \vu\vu^* \right)$ is the orthogonal projection onto $W_{\Sigma} = \vu^{\perp}$. 
Likewise, to evaluate $\mH^*\hat\vq'$ we solve the linear system
\begin{equation}
\label{eqn: linSys_proj2}
    \mT^*\hat\vw' = \hat\vq',\quad \hat\vh' = \left(\mI - \vu\vu^* \right)\hat\vw'.
\end{equation}
Since $\mT$ is often singular or poorly conditioned, solving (\ref{eqn: linSys_proj}) or (\ref{eqn: linSys_proj2}) may be problematic.
If $\mT$ is exactly singular, both direct solution algorithms and iterative solvers will fail.
If $\mT$ is very poorly conditioned, exact solvers based on matrix decompositions (e.g., LU decomposition) may be a viable option in small- and moderate-sized problems. 
As the size of the problem increases, the cost associated with performing a matrix decomposition grows polynomially and exact solution algorithms may become inaccessible. 
This limit can be quickly approached in two-dimensional fluid flows, where $\mT$ may have dimensions on the order $O\left(10^7 \right)-O\left(10^8 \right)$.
Finally, iterative solvers may suffer even in the presence of carefully chosen preconditioners if $\mT$ is poorly conditioned. 
Therefore, if the (near) singularity in $\mT$ could be removed, this could significantly reduce the computational cost required to perform the harmonic resolvent analysis. 


\section{Removing the singularity}
\label{sec: remove_sing}
We now present a computationally efficient way to remove the singularity from $\mT$. 
Throughout this section, we will work with an augmented linear operator 
\begin{equation}
\label{eqn: Ttilde}
    \Ttilde = \begin{bmatrix}
    \mT & \vu \\ \vv^* & 0
    \end{bmatrix},
\end{equation}
where $\vu$ and~$\vv$ are defined as in the previous section: that is, $\vv$ is the unit-norm vector in the direction of the phase shift, and $\vu$ is a unit-norm vector orthogonal to the range of $\mT\vert_\Sigma$ (where $\Sigma = \vv^\perp$).
While $\vv$ is easily computed from the time derivative of the base flow, details on the computation of $\vu$ will be provided in section \ref{subsec: compute_u}. 
We note in passing that operators of the form of (\ref{eqn: Ttilde}) arise in Newton-based harmonic balancing methods for the solution of nonlinear systems, where $\mT$ would be the Jacobian matrix at the $k$th iteration and $\vv$ a phase constraint on the $k$th update. 
In presenting the main results of this paper, we consider two scenarios: when $\mT$ is exactly singular and when $\mT$ is nearly singular. 

\subsection{Singular $\mT$}
\label{subsec: singT}
We first consider the case when $\mT$ is exactly singular, with its one-dimensional nullspace spanned by the direction of phase shift about the base flow. This scenario is likely to arise when the base flow is computed with accuracy close to machine precision using harmonic balancing methods. These methods are ubiquitous in most branches of physics and a similar approach has recently been adopted in \cite{rigas2020nonlinear} to compute a time-periodic and spanwise-periodic solution for the transition to turbulence in a forced boundary layer. 
The main result of this subsection is presented in the proposition below, which states that augmenting $\mT$ as in~\eqref{eqn: Ttilde} removes the singularity, and that the harmonic resolvent operator can be defined in terms of the augmented matrix. 

\begin{proposition}
\label{prop: singT}
Consider the singular value decomposition of $\mT \in \mathbb{C}^{N\times N}$ given by 
\begin{equation}
\label{eqn: svdT}
    \mT = \sum_{j = 1}^{N}\sigma_j \vu_j \vv_j^*,
\end{equation}
where $\sigma_N = 0$, and $\sigma_j>0$ for all $j<N$.
Let $\Ttilde \in \mathbb{C}^{(N+1)\times(N+1)}$ be defined as in (\ref{eqn: Ttilde}), with $\vv = \vv_N$ and $\vu = \vu_N$. Then the following hold:
\begin{enumerate}
    \item $\Ttilde$ is invertible and its singular value decomposition is given by 
    \begin{equation}
        \Ttilde = \sum_{j = 1}^{N-1}\sigma_j \widetilde{\vu}_j\widetilde{\vv}^*_j + \sum_{j = N}^{N+1} \widetilde{\vu}_j\widetilde{\vv}^*_j
    \end{equation}
    where
    \begin{equation}
        \tvv_j = \begin{bmatrix}
        \vv_j \\ 0
        \end{bmatrix},
        \quad \tvu_j = \begin{bmatrix}
        \vu_j \\ 0
        \end{bmatrix},
        \quad j \in \{ 1,2,\cdots,N-1 \}
        \label{eq:SVD_vecs}
    \end{equation}
    and 
    \begin{equation}
        \tvv_N = \begin{bmatrix}
        \vv_N \\ 0
        \end{bmatrix},\quad \tvu_N = \begin{bmatrix}
        0 \\ 1
        \end{bmatrix},
        \quad \tvv_{N+1} = \begin{bmatrix}
        0 \\ 1
        \end{bmatrix},\quad \tvu_{N+1} = \begin{bmatrix}
        \vu_N \\ 0
        \end{bmatrix}.
    \end{equation}
    
    \item The harmonic resolvent, defined in (\ref{eqn: harmonic_res}), is given by
    \begin{equation}
    \label{eqn: harm_res_Ttilde}
        \mH = \sum_{j = 1}^{N-1}\frac{1}{\sigma_j} \vv_j \vu^*_j.
    \end{equation}
    Therefore, the singular values and singular vectors of~$\mH$ can be found from the SVD of~$\Ttilde$, according to~(\ref{eq:SVD_vecs}).
\end{enumerate}
\end{proposition}


The first part of the proposition states that if $\vv$ and $\vu$ in (\ref{eqn: Ttilde}) are properly chosen, then the augmented operator is invertible and well-conditioned. More precisely, the non-zero singular values of $\mT$ agree with $N-1$ of the singular values of $\Ttilde$, and the zero singular value $\sigma_N$ of $\mT$ is replaced by two singular values with value one.
The second statement says that one can easily compute the singular values and singular vectors of the harmonic resolvent from the SVD of the better-conditioned augmented matrix $\Ttilde$.

In order to prove the first statement it suffices to check that $\Ttilde \widetilde{\vv}_j = \sigma_j \widetilde{\vu}_j$, $\forall j \in \{1,2,\cdots,N+1 \}$. 
The proof of the second statement follows immediately from~(\ref{eqn: svdT}) and the definition of the harmonic resolvent in~(\ref{eqn: harmonic_res}).

\subsection{Nearly singular $\mT$}
\label{subsec: nsingT}
We now consider the case when $\mT$ is not exactly singular. This is usually the case when the base flow is computed via numerical integration of the governing equations, as in \cite{padovan2020}, and numerical errors and truncation errors slightly perturb the null singular value.
The perturbed matrix is then invertible, but it may still be poorly conditioned because of this small singular value, and it may therefore become necessary to remove the near singularity in order to improve the performance of iterative solvers. 
As in the previous section, we would like to show that the augmented matrix $\Ttilde$ is better conditioned than $\mT$, and that the SVD of the harmonic resolvent operator can be computed using $\Ttilde$. 

We start by showing that the SVD of the harmonic resolvent can be computed using $\Ttilde$. First and foremost, we recall that the SVD of the harmonic resolvent is performed numerically by computing matrix-vector products of the form $\mH\hat\vh'$ and $\mH^*\hat\vq'$. As mentioned in section \ref{subsec: comp_proc_chall}, these are usually computed using $\mT$, by solving the linear systems (\ref{eqn: linSys_proj}) and (\ref{eqn: linSys_proj2}). The proposition below states that we can compute these matrix-vector products using the better conditioned matrix $\Ttilde$. 
\begin{proposition}
\label{prop: nsingT_equivalence}
Let $\Ttilde$, $\vu$, and $\vv$ be defined as in (\ref{eqn: Ttilde}), with $\mT$ full rank. 
Then the following two statements hold:
\begin{enumerate}
    \item If $\mH$ denotes the harmonic resolvent defined by (\ref{eqn: harmonic_res}), then $\hat\vq' = \mH\hat\vh'$ solves either of the following systems:
    \begin{equation}
            \label{eqn: aug1}
            \begin{bmatrix}
            \mT & \vu \\ \vv^* & 0
            \end{bmatrix}
            \begin{bmatrix}
            \hat\vq' \\ \lambda
            \end{bmatrix} = 
            \begin{bmatrix}
            \hat\vh' \\ 0
            \end{bmatrix} \quad \iff \quad \mT \hat\vq' = \left(\mI - \vu\vu^* \right)\hat\vh'.
    \end{equation}
    \item If $\mH^*$ denotes the adjoint of $\mH$, then $\hat\vh'=\mH^*\hat\vq'$ solves either of the following systems:
    \begin{equation}
    \label{eqn: aug2}
        \begin{bmatrix}
        \mT^* & \vv \\ \vu^* & 0
        \end{bmatrix}
        \begin{bmatrix}
        \hat\vh' \\ \lambda
        \end{bmatrix} = 
        \begin{bmatrix}
        \hat\vq' \\ 0
        \end{bmatrix} \quad \iff \quad \mT^* \hat\vw' = \hat\vq', \quad \hat\vh' = \left(\mI - \vu\vu^* \right) \hat\vw'.
    \end{equation}
\end{enumerate}
\end{proposition}

This means that the action of $\mH$ or $\mH^*$ on a vector (and hence the singular value decomposition of $\mH$) may be computed using either $\mT$ or~$\Ttilde$. The proof of this proposition is given in the appendix. The next proposition establishes that the augmented matrix $\Ttilde$ is better conditioned than the original matrix $\mT$, and we provide a lower bound for the smallest singular value of $\Ttilde$.
\begin{proposition}
\label{prop: nsingT_condition_props}
Let $\sigma$ be the smallest singular value of the operator $\mT\vert_{\Sigma}$ defined in (\ref{eqn: restricted_op}), and suppose that $\norm{\mT\vv} = \varepsilon < 1$. Then we have 
\begin{equation}
\label{eqn: bound_Ttilde}
    \norm{\Ttilde \hat\vz'} \geq \gamma \norm{\hat\vz'},\quad \forall \hat\vz'
\end{equation}
where 
\begin{equation}
\label{eqn: b}
    \gamma = \min\{1-\varepsilon,\sigma\left(1-\varepsilon \right)^{1/2} \}.
\end{equation}
That is, the minimum singular value of $\Ttilde$ is at least $\gamma$. 
\end{proposition}

In most examples we have encountered, the smallest singular value of $\mT$ is in the direction of the phase shift; that is, $\varepsilon$ is smaller than $\sigma$.  In this case, the proposition states that the smallest singular value of $\Ttilde$ is either close to one, or close to~$\sigma$.
Thus, $\Ttilde$ is better conditioned than $\mT$ since we have removed its smallest singular value. The proof of this proposition is also available in the appendix.

\subsection{Computing the vector $\vu$}
\label{subsec: compute_u}
In the previous subsections, we have defined an augmented operator $\Ttilde$ relying on the knowledge of the appropriate vectors $\vv$ and $\vu$ that would remove the singularity from $\mT$.
It is straightforward to compute the vector $\vv$, the unit-norm vector in the direction of the phase shift, which is given by the Fourier coefficients of the time derivative of the base flow.  However, computing $\vu$ requires some care.

Recall that $\vu$ is the orthogonal complement of the range of $\mT\lvert_{\Sigma}$, where $\Sigma=\vv^\perp$, and consider the system 
\begin{equation}
\label{eqn: compute_u}
    \begin{bmatrix}
    \mT^* & \vv \\ \vw^* & 0
    \end{bmatrix}
    \begin{bmatrix}
    \hat\vz' \\ \lambda
    \end{bmatrix} = 
    \begin{bmatrix}
    0 \\ 1
    \end{bmatrix},
\end{equation}
where $\vw$ is an arbitrarily chosen vector. We readily see that the system above can be written as 
\begin{align}
    \mT^*\hat\vz' &= -\lambda \vv \label{eqn: solve_for_z}\\
    \vw^*\hat\vz' &= 1.
\end{align}
Any vector $\hat\vz'$ that satisfies the above equations must be orthogonal to the range of $\mT\vert_{\Sigma}$, since for any $\hat\vq'$ that lies in $\Sigma = \vv^{\perp}$ we have 
\begin{equation}
    \langle \mT\hat\vq',\hat\vz' \rangle = \langle \hat\vq', \mT^*\hat\vz' \rangle = \langle \hat\vq',-\lambda \vv \rangle = 0.
\end{equation}
We therefore solve (\ref{eqn: compute_u}) for $\hat\vz'$ and set $\vu = \hat\vz'/\lVert\hat\vz' \rVert$. 
This approach may fail if we mistakenly choose $\vw$ to lie entirely in the range of $\mT\vert_{\Sigma}$, in which case $\vw^*\hat\vz' = 0$ and the system would have no solution.
This risk, however small it may be, can be avoided by generating $\vw$ in such a way that it is close to the orthogonal complement of the range of $\mT\vert_{\Sigma}$. 
For instance, letting $\mT_{D}$ be the block diagonal components of~$\mT$, one can do so by letting $\vw$ be the solution of the system
\begin{equation}
\label{eqn: prec_w}
    \mT_{D}^*\vw = \vv,
\end{equation}
which can be understood as the block-Jacobi solution of (\ref{eqn: solve_for_z}) with $\lambda = -1$. Even if $\mT$ is exactly singular, $\mT_{D}$ is full-rank and the reader may recognize in this block-diagonal operator the linear operator that governs the linearized dynamics about the temporal mean (i.e., its inverse would contain the well-known resolvent operators discussed in \cite{MRJ2005,BJM2010}).

\section{Application to the Kuramoto-Sivashinsky equation}
\label{sec:KS}
We now illustrate the main results on the Kuramoto-Sivashinsky equation, which is a one-dimensional partial differential equation that arises in the description of instabilities on interfaces and flame fronts.  
This equation was chosen for three reasons. First, it exhibits complex spatio-temporal dynamics similar to those that can arise from the Navier-Stokes equation \cite{holmes}. Second, its spatio-temporal discretization is low-dimensional enough that we can compute the entire spectrum of the operators $\mT$ and $\Ttilde$, thereby providing empirical evidence for the theoretical results developed in sections \ref{subsec: singT} and \ref{subsec: nsingT}. 
Third, its spatio-temporal discretization is high-dimensional enough that we can demonstrate the faster convergence of Krylov solvers when the phase-shift singularity is removed. 

We consider the equation in the form
\begin{equation}
\label{eqn: KS}
    \frac{\partial u}{\partial t} = - u\frac{\partial u}{\partial x} -\frac{\partial^2 u}{\partial x^2} - \frac{\partial^4 u}{\partial x^4},
\end{equation}
where the state $u(x,t)$ is defined over the periodic spatial domain $\mathcal{X} = \left[-L/2,L/2 \right]$. Throughout this section, we consider a domain of length $L = 39$, for which there exist a chaotic attractor and a number of unstable time-periodic orbits \cite{lasagna}. We specify, for the sake of completeness, that the spatial discretization is performed using a Fourier-spectral method, and we retained $32$ spatial wavenumbers.

\begin{figure}
\centering 
\subfloat{
\begin{tikzonimage}[trim = 220 30 60 30, clip=true,width=0.255\textwidth]{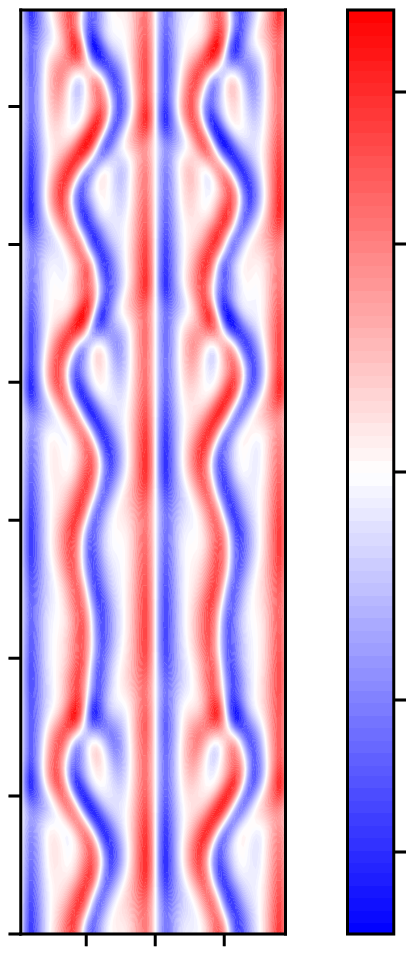}
\foreach \y [count =\i] [evaluate=\y using 0.17+0.14*(\i-1), evaluate=\y as \k using int((\i)*20)] in {0,1,...,5}
{\node[left] at (0.25,\y) {$\k$};}
\node[above, rotate=90] at (0.07, 0.5) {$t$};
\node at (0.34,-0.02) {$-10$};
\node at (0.477,-0.02) {$0$};
\node at (0.595,-0.02) {$10$};
\node[below] at (0.477,-0.04) {$x$};
\node at (0.98,0.1) {$-2.5$};
\node at (0.98,0.255) {$-1.5$};
\node at (0.96,0.485) {$0$};
\node at (0.98,0.725) {$1.5$};
\node at (0.98,0.875) {$2.5$};
\node[rotate=270] at (1.1,0.5) {$U(x,t)$};
\node at (0.6,0.9) {\textit{(a)}};
\end{tikzonimage}}
\hspace{2ex}
\subfloat{
\begin{tikzonimage}[trim = 1 30 30 30, clip=true,width=0.58\textwidth]{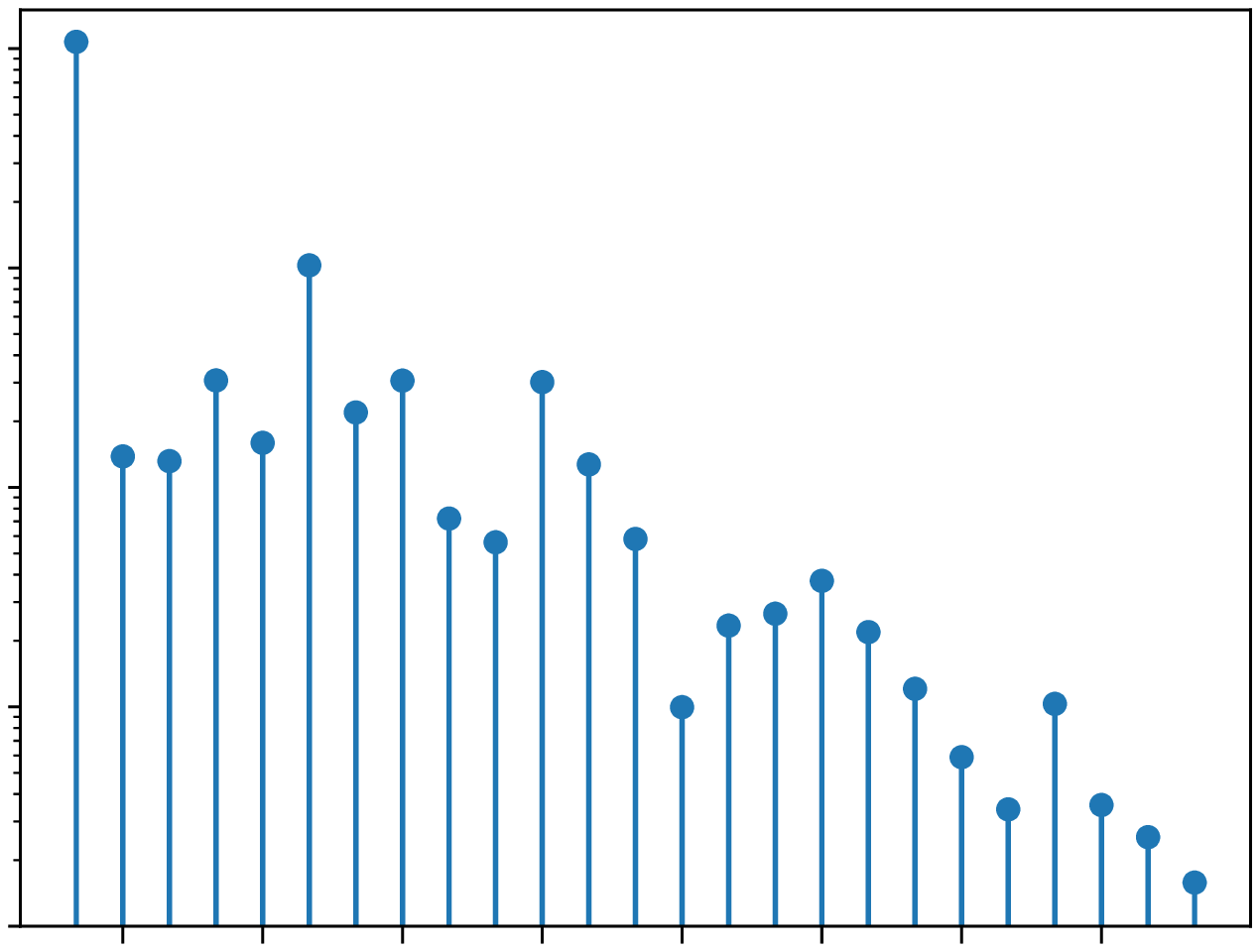}
\node at (0.08,0.04) {$10^{-4}$};
\node at (0.08,0.26) {$10^{-3}$};
\node at (0.08,0.48) {$10^{-2}$};
\node at (0.08,0.7) {$10^{-1}$};
\node at (0.08,0.93) {$10^{0}$};
\node[above, rotate=90] at (0.02, 0.5) {$\langle \hat{U}_{k\omega_f} , \hat{U}_{k\omega_f}\rangle$};
\foreach \x [count =\i] [evaluate=\x using 0.205+0.094*(\i-1), evaluate=\x as \k using int(\i + (\i-1)*2)] in {0,1,...,7}
{\node at (\x,-0.025) {$\k$};}
\node at (0.98,-0.025) {$\times \omega_f$};
\node at (0.55,-0.1) {$k \omega_f$};
\node at (0.9,0.9) {\textit{(b)}};
\end{tikzonimage}}
\caption{We show~\textit{(a)} one of the unstable periodic orbits, with period~$T = 2\pi/\omega_f \approx 134.9$, computed using an harmonic balancing approach, and~\textit{(b)} its energy spectrum.}
\label{fig: KS_snapshots}
\end{figure}

We henceforth omit the spatial dependence of the state variable for notational simplicity. Given a periodic orbit, denoted $U(t)$, we linearize the dynamics by performing the following expansion of the state
\begin{equation}
\label{eqn: decomp_u}
    u(t) = U(t) + u'(t) = \sum_{\omega \in \Omega_b} \hat{U}_{\omega}e^{i\omega t} + \sum_{\omega \in \Omega} \hat{u}'_{\omega}e^{i\omega t},
\end{equation}
where $u'(t)$ are time periodic perturbations about the periodic orbit. As discussed in section \ref{sec: math_formulation}, $\Omega_b$ is the set of frequencies associated with the periodic base flow, while $\Omega$ is the set of frequencies associated with the perturbations. 
We henceforth take $\Omega = \{-24,-23,\cdots,24 \}\omega_f$, where $\omega_f$ is the fundamental frequency of oscillation. 
Upon substituting (\ref{eqn: decomp_u}) into the nonlinear dynamics given by (\ref{eqn: KS}), we obtain 
\begin{equation}
\label{eqn: lin_sys}
    i\omega \hat{u}'_{\omega} = -\left(\frac{\partial^2}{\partial x^2} + \frac{\partial^4}{\partial x^4} \right)\hat{u}'_{\omega} - \sum_{\alpha \in \Omega} \left[\hat{U}_{\omega-\alpha}\frac{\partial}{\partial x} + \frac{\partial \hat{U}_{\omega-\alpha}}{\partial x}\right]\hat{u}'_{\alpha} + \hat{h}'_{\omega},\quad \forall \omega \in \Omega
\end{equation}
where $\hat{h}'_{\omega}$ contains all the nonlinear terms at frequency $\omega$. 
Letting $\hat{u}'$ denote the collection of all Fourier modes, formula (\ref{eqn: lin_sys}) can be written compactly as
\begin{equation}
    \mT \hat{u}' = \hat{h}'.
\end{equation}
We can now verify the theoretical results obtained in section \ref{sec: remove_sing}. Throughout the remainder of this section, the augmented operator $\Ttilde$ is defined as in (\ref{eqn: Ttilde}), where $\vv$ is the unit norm vector in the direction of a phase shift given by the time derivative of $U(t)$, and $\vu$ is computed following the procedure described in section \ref{subsec: compute_u}.

\subsection{Singular $\mT$}
As discussed in the previous sections, when the base flow $U(t)$ satisfies the dynamics exactly, the matrix $\mT$ will be singular with a one-dimensional nullspace in the direction of the phase shift about the base flow. 
We compute one of the unstable periodic orbits that exist for the chosen configuration of the Kuramoto-Sivashinsky equation using a harmonic balancing method. In the Fourier expansion of the candidate solution we consider 24 harmonics of the (yet unknown) fundamental frequency, and we therefore let $\Omega_b = \{-24,-23,\cdots,24 \}\omega_f$.
The spatio-temporal evolution and the energy spectrum of this orbit are shown in figures \ref{fig: KS_snapshots}a and \ref{fig: KS_snapshots}b, and the fundamental period of oscillation is found to be $T = 2\pi/\omega_f \approx 134.9$. 

\begin{figure}
\vspace{-5ex}
\hspace{-10ex}
\begin{minipage}{0.4\textwidth}
\begin{tikzonimage}[trim = 20 30 30 30, clip=true,width=1.05\textwidth]{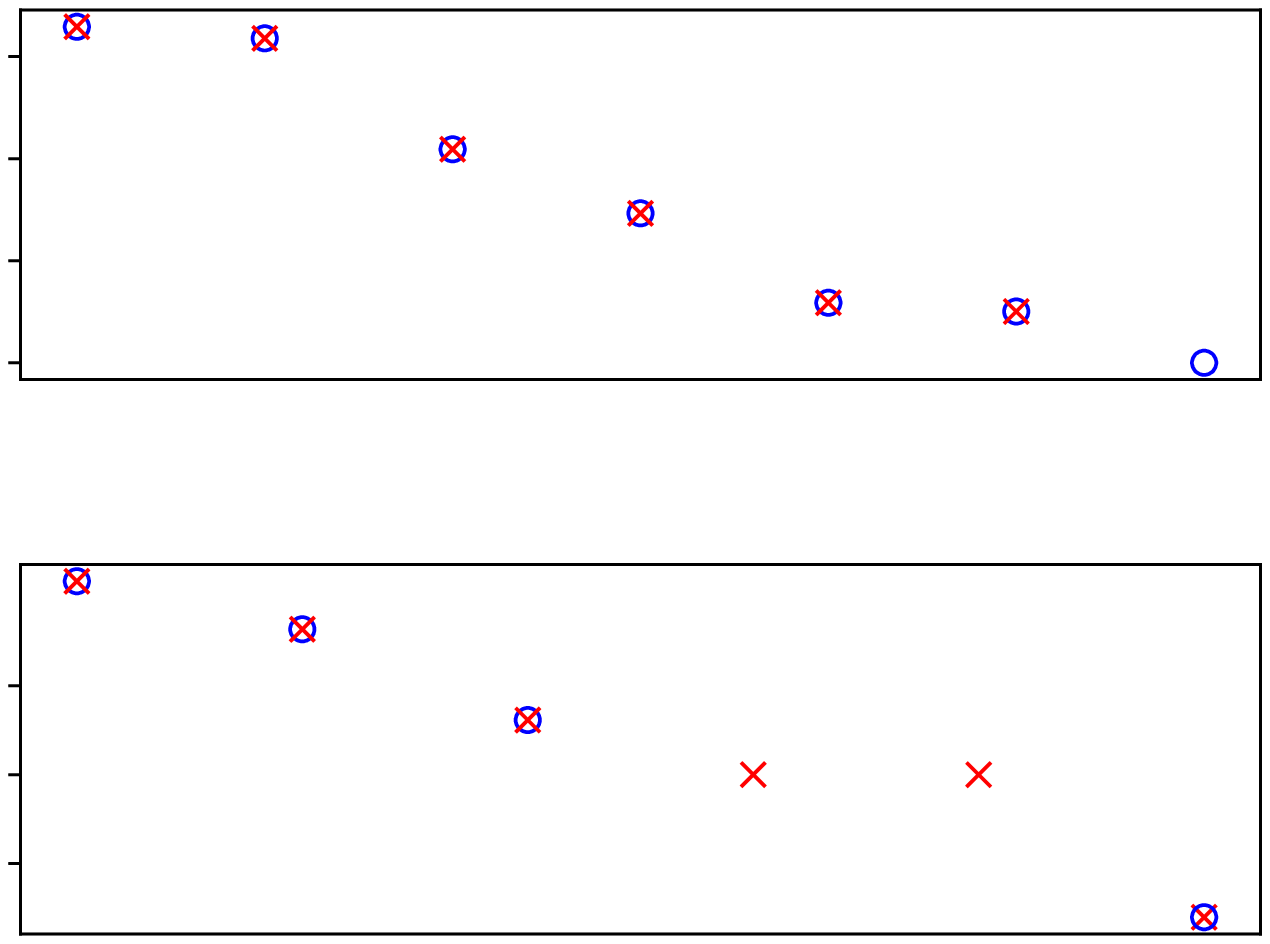}
\node at (0.05,0.605) {$0$};
\node at (0.01,0.71) {$0.002$};
\node at (0.01,0.81) {$0.004$};
\node at (0.01,0.915) {$0.006$};
\node at (0.01,0.19) {$1.000$};
\node at (0.01,0.28) {$1.002$};
\node at (0.01,0.1) {$0.998$};
\draw [dashed] (0.11,0.189) -- (0.953,0.189);
\node at (0.52,1.015) {Sing. values of $\mT$ and $\Ttilde$ near zero};
\node at (0.52,0.46) {Sing. values of $\mT$ and $\Ttilde$ near unity};
\node at (0.92,0.645) {$\sigma_{N}$};
\node at (0.9,0.9) {\textit{(a)}};
\node at (0.9,0.34) {\textit{(b)}};
\end{tikzonimage}
\end{minipage}
\hspace{6ex}
\begin{minipage}{0.4\textwidth}
\vspace{7ex}
\begin{tikzonimage}[trim = 20 30 30 30, clip=true,width=1.05\textwidth]{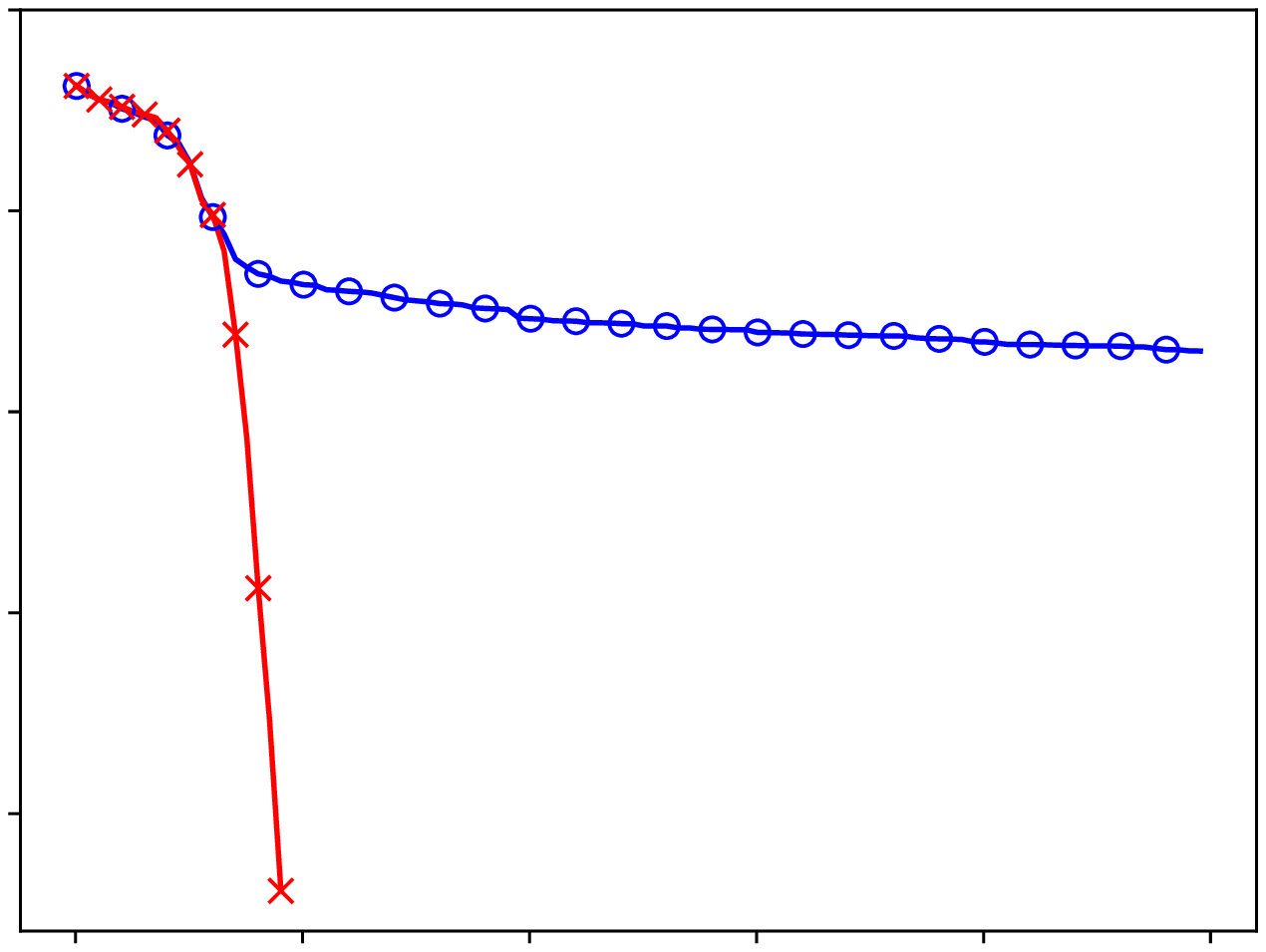}
\node at (0.02,0.15) {$10^{-8}$};
\node at (0.02,0.35) {$10^{-6}$};
\node at (0.02,0.55) {$10^{-4}$};
\node at (0.02,0.75) {$10^{-2}$};
\node at (0.02,0.96) {$10^{0}$};
\node at (0.135,-0.03) {$0$};
\node at (0.29,-0.03) {$200$};
\node at (0.45,-0.03) {$400$};
\node at (0.61,-0.03) {$600$};
\node at (0.77,-0.03) {$800$};
\node at (0.93,-0.03) {$1000$};

\node at (0.5,-0.13) {GMRES iteration};
\node[rotate=90] at (-0.1,0.5) {$\lVert \mathrm{Residual} \rVert_2$};

\node at (0.9,0.9) {\textit{(c)}};
\end{tikzonimage}
\end{minipage}
\caption{We show~\textit{(a)} the singular values of $\mT$ and $\Ttilde$ close to zero (the smallest singular value of $\mT$ is $\sigma_N \sim O(10^{-13})$, indicating that $\mT$ is singular for practical purposes), ~\textit{(b)} the singular values of $\mT$ and $\Ttilde$ close to unity and ~\textit{(c)} a convergence plot obtained by solving the linear systems in (\ref{eqn: conv_eq}) and (\ref{eqn: conv_eq2}). The dashed line in \textit{(b)} is an extension of the 1-tick on the y axis. (\emph{$\circ$: data for $\mT$; $\times$: data for $\Ttilde$}).}
\label{fig: KS_sing_fig}
\end{figure}

In figures \ref{fig: KS_sing_fig}a and \ref{fig: KS_sing_fig}b we show the relevant singular values of $\mT$ and~$\Ttilde$. 
Specifically, from figure \ref{fig: KS_sing_fig}a, we observe that while $\mT$ has a zero singular value ($\sigma_N \approx 2.6\times 10^{-13}$), $\Ttilde$ does not, and its smallest singular value agrees with the smallest non-zero singular value of $\mT$. 
Furthermore, we observe from figure \ref{fig: KS_sing_fig}b that two singular values of $\Ttilde$ have value one. 
Finally, we notice that except for the highlighted differences, the singular values of $\mT$ and $\Ttilde$ agree exactly.  
These observations are as we expect from P
proposition~\ref{prop: singT}.

Figure \ref{fig: KS_sing_fig}c shows a representative convergence plot illustrating the performance of the block-Jacobi-preconditioned GMRES solver on the linear systems
\begin{align}
    \mT \vx &= \vh \label{eqn: conv_eq}\\ 
    \Ttilde\widetilde{\vx} &= \left(\vh,0 \right), \label{eqn: conv_eq2}
\end{align}
where $\vh$ is a random unit-norm vector. 
The benefits of removing the singularity from the linear operator $\mT$ are clear. 
Specifically, we see that the solver using equation (\ref{eqn: conv_eq}) (blue curve) plateaus at a residual on the order of $10^{-3}$, while the solver using equation (\ref{eqn: conv_eq2}) (red curve) converges to the desired tolerance in fewer than 200 iterations. 

\subsection{Nearly singular $\mT$}

We now consider the case when the base flow $U(t)$ does not satisfy the governing equation exactly. 
We introduce an error in the base flow described in the previous section by truncating its highest frequency component, so that $\Omega_b = \{-23,-22,\cdots,23\}\omega_f$. The set of frequencies $\Omega$ associated with the perturbed state $u'(t)$ is kept unchanged. 

\begin{figure}
\vspace{-5ex}
\hspace{-10ex}
\begin{minipage}{0.4\textwidth}
\centering
\begin{tikzonimage}[trim = 20 30 30 30, clip=true,width=1.05\textwidth]{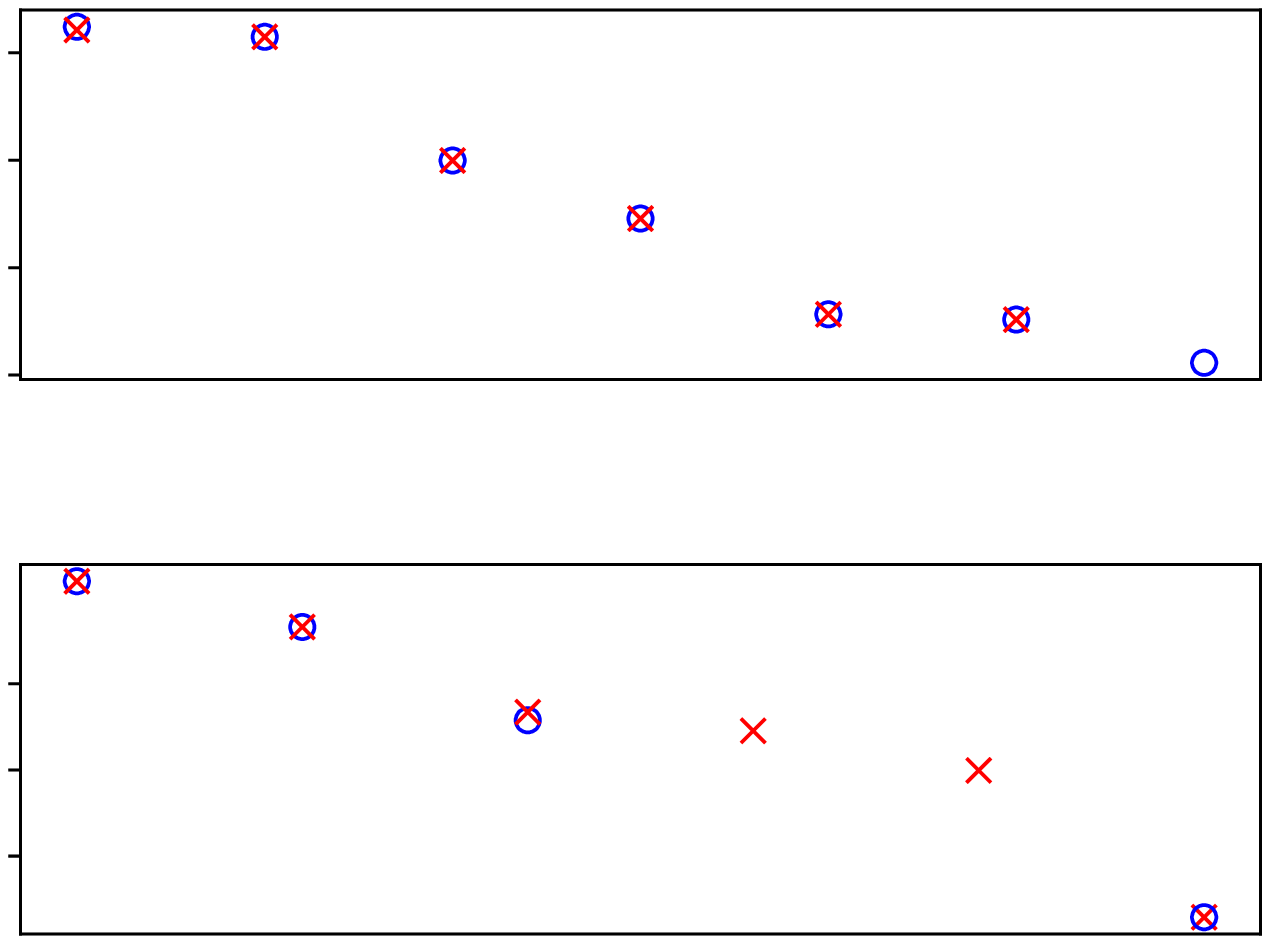}
\node at (0.05,0.605) {$0$};
\node at (0.01,0.71) {$0.002$};
\node at (0.01,0.81) {$0.004$};
\node at (0.01,0.915) {$0.006$};
\node at (0.01,0.19) {$1.000$};
\node at (0.01,0.28) {$1.002$};
\node at (0.01,0.1) {$0.998$};
\draw [dashed] (0.11,0.195) -- (0.953,0.195);
\node at (0.52,1.015) {Sing. values of $\mT$ and $\Ttilde$ near zero};
\node at (0.52,0.46) {Sing. values of $\mT$ and $\Ttilde$ near unity};
\node at (0.92,0.645) {$\sigma_{N}$};
\node at (0.9,0.9) {\textit{(a)}};
\node at (0.9,0.34) {\textit{(b)}};
\end{tikzonimage}
\end{minipage}
\hspace{6ex}
\begin{minipage}{0.4\textwidth}
\centering
\vspace{7ex}
\begin{tikzonimage}[trim = 20 30 30 30, clip=true,width=1.05\textwidth]{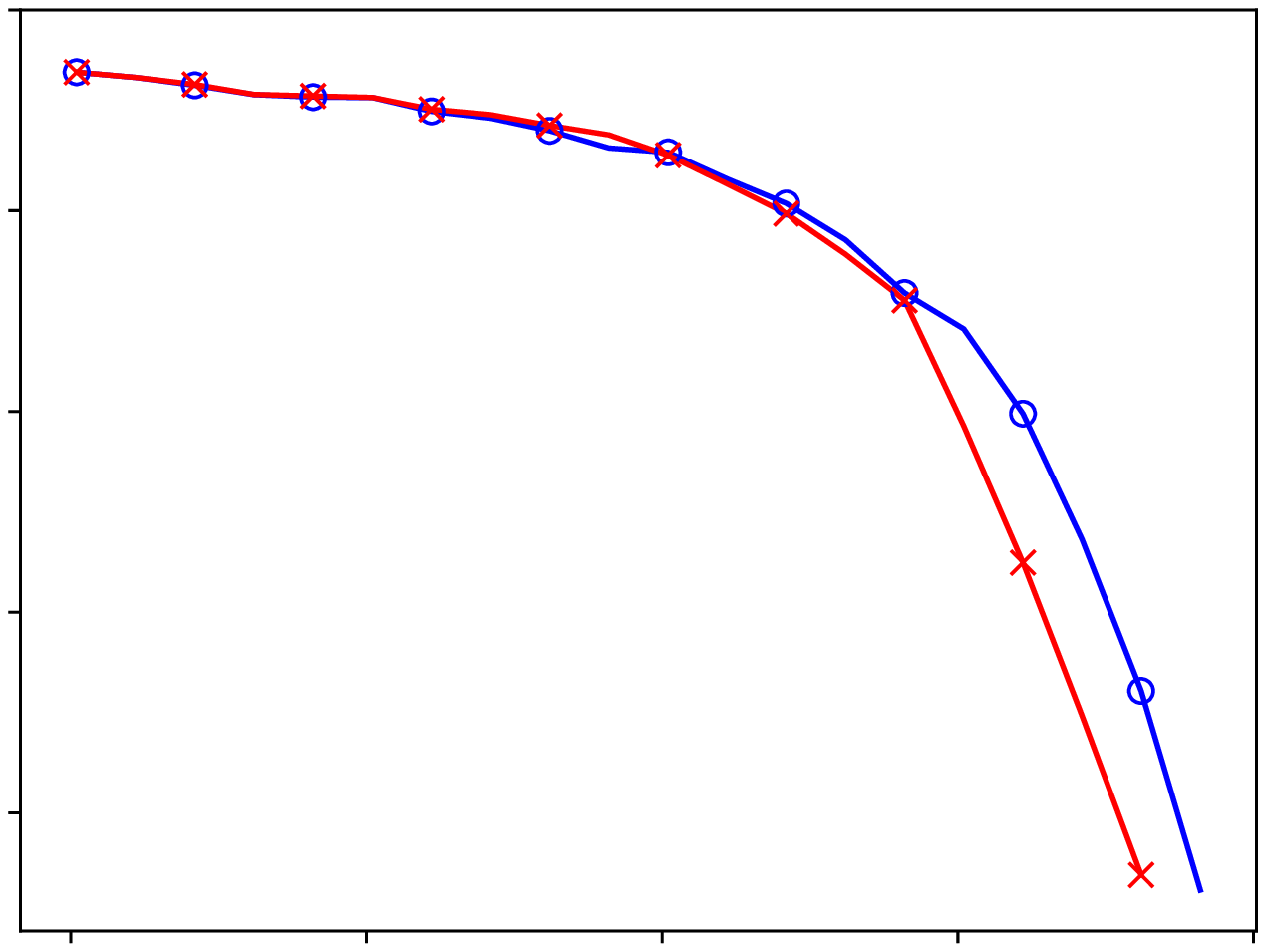}
\node at (0.02,0.15) {$10^{-8}$};
\node at (0.02,0.35) {$10^{-6}$};
\node at (0.02,0.55) {$10^{-4}$};
\node at (0.02,0.75) {$10^{-2}$};
\node at (0.02,0.96) {$10^{0}$};
\node at (0.13,-0.03) {$0$};
\node at (0.34,-0.03) {$50$};
\node at (0.54,-0.03) {$100$};
\node at (0.75,-0.03) {$150$};
\node at (0.96,-0.03) {$200$};

\node at (0.5,-0.13) {GMRES iteration};
\node[rotate=90] at (-0.1,0.5) {$\lVert \mathrm{Residual} \rVert_2$};

\node at (0.9,0.9) {\textit{(c)}};
\end{tikzonimage}
\end{minipage}
\caption{Analog of figure \ref{fig: KS_sing_fig}, showing~\textit{(a)} the singular values of $\mT$ and $\Ttilde$ close to zero (the smallest singular value of $\mT$ is $\sigma_N \sim O(10^{-4})$), ~\textit{(b)} the singular values of $\mT$ and $\Ttilde$ close to unity and ~\textit{(c)} a convergence plot obtained by solving the linear systems in (\ref{eqn: conv_eq}) and (\ref{eqn: conv_eq2}). The dashed line in \textit{(b)} is an extension of the 1-tick on the y axis. (\emph{$\circ$: data for $\mT$; $\times$: data for $\Ttilde$}).}
\label{fig: KS_nsing_fig}
\end{figure}

As expected, $\mT$ is now non-singular, and its smallest singular value (shown in figure \ref{fig: KS_nsing_fig}a) has order of magnitude $O(10^{-4})$. 
Constructing $\Ttilde$ is beneficial nonetheless, and we see from figure \ref{fig: KS_nsing_fig}a that the smallest singular value is removed, and it is replaced with two singular values with value $\approx 1$ (see figure \ref{fig: KS_nsing_fig}b). 
We see, furthermore, that removing the near singularity has introduced a slight perturbation in the spectrum, since the remaining singular values of $\Ttilde$ do not agree exactly with those of $\mT$. 
We recall, however, that we are ultimately interested in the singular values and singular vectors of the harmonic resolvent $\mH$, and we have shown via proposition \ref{prop: nsingT_equivalence} that we are allowed to use $\Ttilde$ to compute its singular value decomposition. 

Finally, figure \ref{fig: KS_nsing_fig}c shows a representative convergence plot illustrating the performance of the block-Jacobi-preconditioned GMRES solver on the linear systems in equations (\ref{eqn: conv_eq}) and (\ref{eqn: conv_eq2}). 
Although the speed-up in convergence is not as substantial as the one shown in figure \ref{fig: KS_sing_fig}c, we observe that removing the smallest singular value still leads to slightly faster convergence. 

\vspace{-1ex}
\section{Conclusion}
It is well-known that linear time-periodic dynamics are neutrally stable in the direction of a phase shift about the time-periodic orbit, and this property manifests itself in the form of a singularity in the linear operator that governs the dynamics. 
This singularity, in turn, leads to numerical difficulties in the context of the harmonic resolvent analysis, where the computation of the singular value decomposition of the harmonic resolvent operator requires the inversion of this singular operator. 

We have proposed a computationally inexpensive solution to this problem, showing that a suitable augmentation of the singular matrix leads to the removal of the singularity and to a significant improvement in the condition properties of the resulting augmented linear operator. 
In our discussion, we have considered the cases when the operator is exactly singular and when the operator is nearly singular. We then used the Kuramoto-Sivashinsky equation as an example to demonstrate that in both cases it is convenient to remove the (near) singularity in order to improve the convergence properties of classical Krylov-based linear solvers.

\section*{Acknowledgements}
The authors wish to thank Samuel Otto for providing useful comments that helped
improve the presentation of this paper. This material is based upon work supported by the Air Force Office of Scientific Research under award number FA9550-17-1-0084.

\section*{Conflict of interest}
The authors declare that they have no conflict of interest.

\appendix

\section{Proof of proposition 2}
We start by verifying the first statement, which we copy here for clarity. 
\begin{equation}
    \begin{bmatrix}
    \mT & \vu \\ \vv^* & 0
    \end{bmatrix}
    \begin{bmatrix}
    \hat\vq' \\ \lambda
    \end{bmatrix} = 
    \begin{bmatrix}
    \hat\vh' \\ 0
    \end{bmatrix} \quad \iff \quad \mT \hat\vq' = \left(\mI - \vu\vu^* \right)\hat\vh'.
\end{equation}
We first recall that $\vv$ is the unit-norm orthogonal complement of $\Sigma$, while $\vu$ is the unit-norm orthogonal complement of the range of $\mT\vert_{\Sigma}$.
From the second row of the augmented linear system we see that $\langle \hat\vq',\vv\rangle = 0$, which implies that $\hat\vq' \in \Sigma$, and hence $\langle \mT\hat\vq', \vu \rangle = 0$. 
We proceed by taking the inner product of the first row of the augmented system with $\vu$, to obtain
\begin{equation}
    \langle \mT\hat\vq' ,\vu \rangle + \lambda \langle \vu,\vu \rangle = \langle \hat\vh',\vu \rangle \quad \Rightarrow \quad \lambda = \langle \hat\vh',\vu \rangle,
\end{equation}
where we have used $\langle \vu,\vu \rangle = 1$. Substituting $\lambda$ into the first row of the augmented system establishes the desired result. 

We proceed by verifying the second statement of the proposition, given by 
\begin{equation}
\label{eqn: equiv2}
    \begin{bmatrix}
    \mT^* & \vv \\ \vu^* & 0
    \end{bmatrix}
    \begin{bmatrix}
    \hat\vh' \\ \lambda
    \end{bmatrix} = 
    \begin{bmatrix}
    \hat\vq' \\ 0
    \end{bmatrix} \quad \iff \quad \mT^* \hat\vw' = \hat\vq', \quad \hat\vh' = \left(\mI - \vu\vu^* \right) \hat\vw'.
\end{equation}
We first leverage the fact that $\mT$ is invertible and we recall, as discussed in section \ref{subsec: compute_u}, that $\vu$ is given by
\begin{equation}
\label{eqn: vz}
    \vz = \left(\mT^* \right)^{-1}\vv,\quad \vu = \vz/\norm{\vz}.
\end{equation}
From the first row of the augmented system and using (\ref{eqn: vz}) we have
\begin{equation}
\label{eqn: vh}
    \hat\vh' = \left(\mT^* \right)^{-1}\left(\hat\vq' - \lambda \vv \right) = \left(\mT^* \right)^{-1}\hat\vq' - \lambda\norm{\vz} \vu.
\end{equation}
By the second row of the augmented system we have $\langle\hat\vh' ,\vu\rangle = 0$, thus, using (\ref{eqn: vh}) we have
\begin{equation}
    \langle\left(\mT^* \right)^{-1}\hat\vq' ,\vu\rangle - \lambda \norm{\vz}\langle \vu,\vu \rangle = 0 \quad \Rightarrow\quad \lambda = \frac{1}{\norm{\vz}}\langle\left(\mT^* \right)^{-1}\hat\vq' ,\vu\rangle.
\end{equation}
Substituting $\lambda$ into (\ref{eqn: vh}) we obtain
\begin{equation}
    \hat\vh' = \left(\mI - \vu\vu^* \right)\left(\mT^* \right)^{-1}\hat\vq',
\end{equation}
which is precisely what is given on the right of the equivalence sign in (\ref{eqn: equiv2}), with $\hat\vw' = \left(\mT^* \right)^{-1}\hat\vq'$.

\section{Proof of proposition 3}
First, observe that since $\sigma$ is the smallest singular value of $\mT\vert_\Sigma$, it follows that
\begin{equation}
    \|\mT \hat\vq'\| \ge \sigma \|\hat\vq'\|,\qquad\text{for all $\hat\vq'\in\Sigma$}.
    \label{eqn: bound_T}
\end{equation}
Let us write $\hat\vz' = (\hat\vq' + \alpha \vv,\lambda)$, where $\hat\vq' \in \Sigma$ and $\alpha,\lambda \in \mathbb{C}$. 
Note that any $\hat\vz'$ can be written this way, since $\Sigma$ is the orthogonal complement of $\vv$. Then
\begin{equation*}
    \norm{\hat\vz'}^2 = \norm{\hat\vq' + \alpha \vv}^2 + \abs{\lambda}^2 = \norm{\hat\vq'}^2 + \abs{\alpha}^2 + \abs{\lambda}^2,
\end{equation*}
since $\vv \perp \hat\vq'$ and $\norm{\vv}=1$. Furthermore, 
\begin{equation*}
    \Ttilde\hat\vz' = 
    \begin{bmatrix}
    \mT(\hat\vq' + \alpha \vv) + \lambda \vu \\ \vv^*(\hat\vq' + \alpha \vv)
    \end{bmatrix} = 
    \begin{bmatrix}
    \mT(\hat\vq' + \alpha \vv) + \lambda \vu \\ \alpha
    \end{bmatrix}
\end{equation*}
again since $\vv \perp \hat\vq'$ and $\norm{\vv}=1$. Now, 

\begin{align*}
    \norm{\Ttilde \hat\vz'}^2 &= \norm{\mT\hat\vq' + \alpha \mT\vv + \lambda \vu}^2 + \abs{\alpha}^2 =\\ 
    &= \norm{\mT\hat\vq'}^2 + \abs{\alpha}^2 \norm{\mT\vv}^2 + \abs{\lambda}^2\norm{\vu}^2 +  \\
    &+ 2 \mathrm{Re}\left[\langle \alpha \mT\vv,\mT\hat\vq' \rangle + \langle \lambda\vu,\mT\hat\vq' \rangle + \langle \alpha\mT\vv,\lambda \vu \rangle\right] + \abs{\alpha}^2 = \\
    &= \norm{\mT\hat\vq'}^2 + (1+\varepsilon^2)\abs{\alpha}^2 + \abs{\lambda}^2 + \\
    &+ 2 \mathrm{Re}\left[\langle \alpha \mT\vv,\mT\hat\vq' \rangle + \langle \alpha\mT\vv,\lambda \vu \rangle\right]
\end{align*}
since $\vu \perp \mT\hat\vq'$ and $\norm{\mT\vv}=\varepsilon$. Note that for any $x,y$, we have 
\begin{equation*}
    -\mathrm{Re}\langle x,y\rangle \leq \abs{\langle x,y\rangle}\leq \norm{x}\cdot \norm{y},
\end{equation*}
thanks to the Cauchy-Schwarz inequality. Therefore, 
\begin{equation*}
    \norm{\Ttilde\vz}^2 \geq \norm{\mT\hat\vq'}^2 + (1+\varepsilon^2)\abs{\alpha}^2 + \abs{\lambda}^2 - 2\varepsilon \abs{\alpha}\norm{\mT\hat\vq'} - 2\varepsilon \abs{\alpha}\abs{\lambda}, 
\end{equation*}
since $\mT\vv = \varepsilon$. Since $-2ab \geq -(a^2 + b^2)$ for any $a,b$, we have 
\begin{align*}
    \norm{\Ttilde\hat\vz'}^2 &\geq \norm{\mT\hat\vq'}^2 + (1+\varepsilon^2)\abs{\alpha}^2 + \abs{\lambda}^2 
    - (\varepsilon\abs{\alpha}^2 + \varepsilon\norm{\mT\hat\vq'}^2) - (\varepsilon\abs{\alpha}^2 + \varepsilon\abs{\lambda}^2)  \\
    &= (1-\varepsilon)\norm{\mT\hat\vq'}^2 + (1-2\varepsilon+\varepsilon^2)\abs{\alpha}^2 + (1-\varepsilon)\abs{\lambda}^2  \\
    &\geq (1-\varepsilon)\sigma^2\norm{\hat\vq'}^2 + (1-\varepsilon)^2\abs{\alpha}^2 + (1-\varepsilon)\abs{\lambda}^2
\end{align*}
thanks to (\ref{eqn: bound_T}), since $0<1-\varepsilon<1$. So if we let 
\begin{equation*}
    \gamma^2 = \min\{(1-\varepsilon)\sigma^2,(1-\varepsilon)^2,(1-\varepsilon) \} = \min\{(1-\varepsilon)\sigma^2,(1-\varepsilon)^2 \},
\end{equation*}
we have 
\begin{equation*}
    \norm{\Ttilde\hat\vz'}^2 \geq \gamma^2(\norm{\hat\vq'}^2 + \abs{\alpha}^2 + \abs{\lambda}^2) = \gamma^2\norm{\hat\vz'}^2.
\end{equation*}
Taking square roots then establishes (\ref{eqn: bound_Ttilde}) for $\gamma$ given by (\ref{eqn: b}).


\bibliographystyle{spphys}       
\bibliography{RowleyGroupReferences}

\end{document}